\documentclass[preprint,showpacs,preprintnumbers,amsmath,amssymb]{revtex4}

\usepackage{graphicx}
\usepackage{dcolumn}
\usepackage{bm}

\begin{document}

\title{On the nature of amorphous polymorphism of water}

\author{M.M.~Koza$^{1}$,
        B.~Geil$^{2}$,
        K.~Winkel$^{2}$,
        C.~K\"ohler$^{2}$,
        F.~Czeschka$^{1,3}$,
        M.~Scheuermann$^{2}$
        H.~Schober$^{1}$,
        T.~Hansen$^{1}$.}
\address{$^1$Institut Laue-Langevin, F-38042 Grenoble Cedex, France.\\
         $^2$Institut f\"ur Festk\"orperphysik, TU--Darmstadt, Germany.\\ 
         $^3$Institut f\"ur Festk\"orperphysik, TU--M\"unchen, Germany. 
        }

\date{\today}

\begin{abstract}
We report elastic and inelastic neutron scattering experiments
on different amorphous ice modifications.
It is shown that an amorphous structure (HDA') indiscernible
from the high--density phase (HDA), obtained by compression
of crystalline ice, can be formed from the very high--density phase (vHDA)
as an intermediate stage of the transition of vHDA into its
low--density modification (LDA').
Both, HDA and HDA' exhibit comparable small angle scattering
signals characterizing them as structures heterogeneous on a length
scale of a few nano--meters.
The homogeneous structures are the initial and final transition stages
vHDA and LDA', respectively. 
Despite, their apparent structural identity on a local scale HDA and HDA'
differ in their transition kinetics explored by {\it in situ} experiments.
The activation energy of the vHDA--to--LDA' transition is at least 
20~kJ/mol higher than the activation energy of the HDA--to--LDA transition. 
\end{abstract}


\maketitle

Amorphous polymorphism is strongly linked to water, 
where for the first time two distinct amorphous modifications, namely a high--density 
amorphous (HDA,  $\rho \approx 39$~mol./nm$^3$) and a low--density 
amorphous (LDA, $\rho \approx 31$~mol./nm$^3$) ice state could be 
prepared \cite{Mishima-Nature-1984}.
The existence of HDA and LDA and the characteristics of the transition between 
these two modifications, often referred to as first--order like, have triggered major 
experimental and theoretical efforts that all aim for a conclusive explanation 
of amorphous polymorphism \cite{Mishima-Nature-1998}.
It has, in particular, been conjectured from Molecular Dynamics (MD) simulations,
that water in the super-cooled region exhibits a second
critical point and a first--order transition line
separating two liquid phases towards lower temperatures
\cite{Poole-PRL-1994}.
In this picture HDA and LDA are supposed to be glassy representatives of these
two liquids.

The scenario of two super-cooled liquid phases has been recently questioned
by the discovery of a third disordered modification apparently distinct from HDA
and LDA and called, due to its higher density
($\rho \approx 41$~mol./nm$^3$), very high--density
amorphous (vHDA) ice \cite{Loerting-PCCP-2001}.
Moreover, latest computer simulations are in apparent agreement
with this finding suggesting the presence
of "multiple liquid--liquid transitions in super-cooled water"
\cite{Brovchenko-JCP-2003}.

Amorphous systems like liquids are isotropic and thus possess
by definition no particular global symmetry. 
In the glassy state they are in addition non-ergodic,
i.e. they are prone to relax their structures following distinct
energetic pathways that may or may not be accessible during the experiment. 
Thus, unless there are clear indications for thermodynamic transitions,
it is not possible to assign distinct phases to an amorphous system.
The evolution of an amorphous system as a function of temperature,
pressure and time may, however, be characterized via the changes encountered
in local structural units. 
These can e.g. be studied with wide-angle diffraction.
If the structural changes are significant this approach allows for
a characterisation of the amorphous states.
We would like to recall that due to the non-ergodic nature of the sample
what we call a state is not necessarily completely characterized
by thermodynamic variables but may well depend on the sample history.
In the case of water the nearest-neighbor coordination number
has been proposed as a criterium  to distinguish
LDA, HDA and vHDA (coordination number of  four, five and six, respectively)
\cite{Finney-PRL-2002a,Finney-PRL-2002b}. 
Another approach recently used in MD simulations bases the analysis
not on the coordination number but on the ring structures
encountered in the system \cite{Martonak-PRL-2004}.
The important question remains whether the local characterization
is sufficient and whether the thus classified states actually 
correspond to distinct phases. 

In this communication we present diffraction and inelastic neutron
scattering experiments on both HDA and vHDA samples.
HDA (D$_2$O) has been formed by slow compression of crystalline
ice I$_{\rm h}$ to 18~kbar at 77~K \cite{Koza-PRL-2000}.
vHDA samples have been obtained by heating HDA (D$_2$O) samples
at high pressure \cite{Loerting-PCCP-2001}.
Here, we present data collected with three vHDA samples formed at
$p=10.5$~kbar $T=145$~K \#~1,
$p=10.5$~kbar $T=155$~K \#~2 and
$p=16$~kbar $T=155$K \#~3.
After the vHDA structure is formed all samples are cooled at
the indicated $p$ to $T=77$~K and retrieved
from the pressure cells, crushed into mm--sized chunks and placed
into proper sample containers.
Please note, that throughout this paper we will refer
to the amorphous ice modifications obtained from vHDA 
by heating as HDA' and LDA'.
We will show that HDA and HDA', a modification of amorphous ice obtained
by heating of vHDA at ambient $p$, are indistinguishable in terms
of local structure and dynamics as probed in reciprocal space
by neutron scattering.
The amount of small angle scattering gives evidence that HDA and HDA'
are both heterogeneous structures. 

Diffraction experiments have been carried out at
the high--flux instrument D20 ($\lambda=2.4$\,\AA\  and 1.3\,\AA )
and inelastic data have been collected at the
time--of--flight spectrometer IN6 ($\lambda=4.1$\,\AA ),
Institut Laue--Langevin, Grenoble, France. 
Standard data corrections for empty container scattering (IN6)
and detector calibration of both instruments,
has been applied.
Elastic data are normalized to the coherent part of
the IN6 response allowing a comparison of all data sets
with each other \cite{data_analysis_01}.
The {\it in situ} transformations of vHDA into LDA' have been
followed in an Helium atmosphere of 200~mbars,
and the corresponding data evaluation have been carried out
according to reference \cite{Koza-JPCM-2003}.
1--2~mbars of Helium pressure have been applied during
the inelastic experiments.

\begin{figure}
\includegraphics[angle=0,width=150mm]{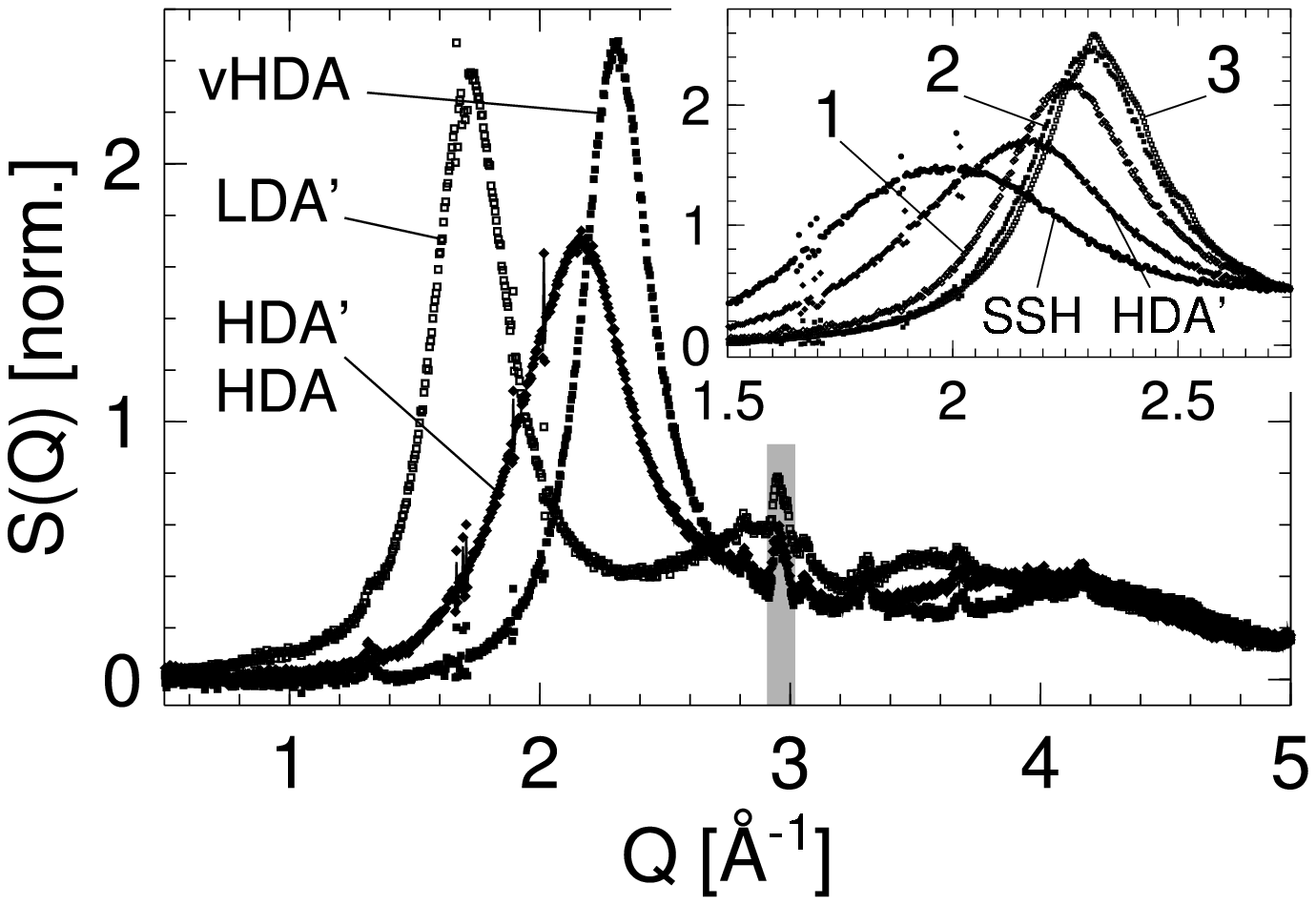}
\caption{Static structure factor $S(Q)$ of 
vHDA \#\,2 ($\blacksquare$), HDA' ($\blacklozenge$)
and LDA' ($\square$), as obtained from vHDA \#\,2,
and of HDA (--), as produced by compressing ice I$_{\rm h}$.
Please note that HDA' and HDA are hardly distinguishable.
The grey shaded area indicates a Bragg--peak
from the sample container.
The insert reports the maximum of $S(Q)$ for vHDA samples
\#\,1, \#\,2 and \#\,3.
Equally shown are HDA' and SSH \#\,2.}
\label{fig_01}
\end{figure}

The static structure factor $S(Q)$ of sample \#\,2 is reported
in Fig.~\ref{fig_01}.
$S(Q)$ of HDA and vHDA is measured at 75\,K, that of HDA' is
obtained during {\it in situ} annealing at 113\,K.
$S(Q)$ of LDA' is determined after annealing at 135\,K.
Fig.~\ref{fig_01} demonstrates that it is possible to select
an intermediate ice modification HDA' such that the $S(Q)$ of HDA'
and HDA coincide.
It furthermore can be shown that to any modification encountered
along the transition from HDA to LDA \cite{Koza-JPCM-2003}
we find a matching partner along the transition from vHDA to LDA'.
In terms of $S(Q)$ the transition states of HDA to LDA appear as
a subset of the transition states of vHDA to LDA'.
Characteristic features of both transitions are the continuous
down-shift of the maximum in $S(Q)$ and the transient broadening
of the intermediate maximum, which shows the largest width in
the middle of the transition \cite{Koza-JPCM-2003,Guthrie-PRB-2003}.
This is demonstrated in the insert of Fig.~\ref{fig_01} by
the $S(Q)$ of vHDA, HDA' and an intermediate state identified as the state
of strongest heterogeneity (SSH), as it is discussed below.

The initial transition stage is a matter of the thermodynamic
conditions, i.e. $p$ and $T$, during sample preparation.
This point is substantiated equally in the insert of Fig.~\ref{fig_01}
where the main peak of $S(Q)$ is plotted for samples 
\#\,1, \#\,2 and \#\,3. 
vHDA of sample \#\,1 appears as an apparent intermediate
state between vHDA and HDA' measured with sample \#\,2.
It can be matched by an intermediate $S(Q)$ of sample \#\,2
transforming into LDA'.
The sample preparation concerning $p$ and $T$ is fully reproducible,
however we cannot exclude that the final state obtained is conditioned
by the history, i.e. by the state of relaxation of the sample
at the time when $p$ and $T$ were applied.

\begin{figure}
\includegraphics[angle=0,width=150mm]{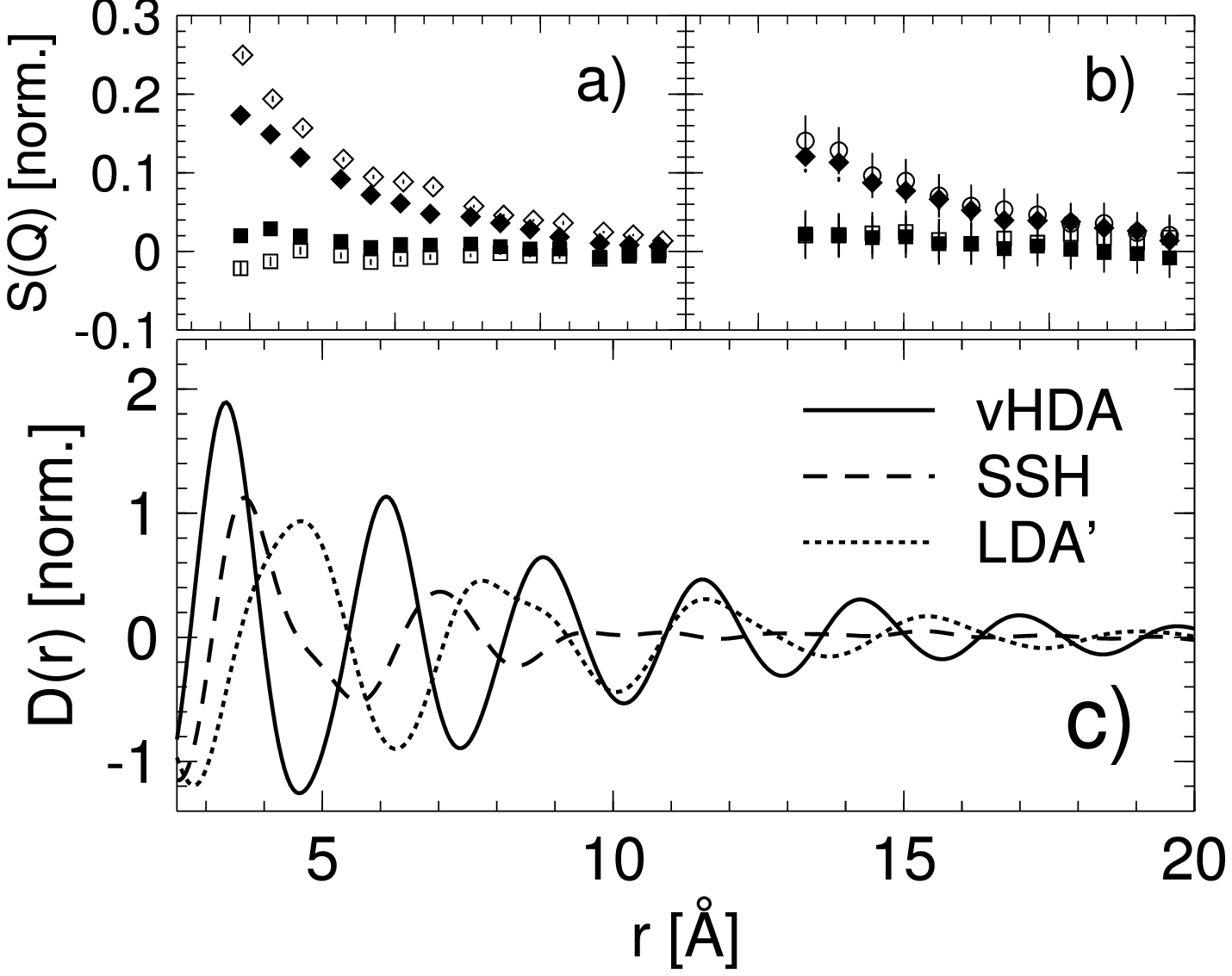}
\caption{(a) $S(Q)$ small angle signal of vHDA ($\blacksquare$), 
LDA' ($\square$), HDA' ($\blacklozenge$) and HDA ($\lozenge$)
measured at IN6 and (b) of vHDA ($\blacksquare$), 
LDA' ($\square$), HDA' ($\blacklozenge$) and SSH ($\circ$)
measured at D20 (\#\,2 in Fig.~\ref{fig_01}).
(c) $D(r)$ calculated for vHDA, LDA' and SSH from
Fig.~\ref{fig_01}.
}
\label{fig_02}
\end{figure}

Another basic characteristic of the transitions is a transient
enhancement of $S(Q)$ at $Q \le 0.6$~\AA $^{-1}$.
Figs.~\ref{fig_02}~a and \ref{fig_02}~b report on the $S(Q)$ determined
with sample \#\,1 in the states vHDA, HDA' and LDA' and
a HDA sample at IN6 and with sample \#\,2 in the states vHDA, HDA', 
LDA' and the state of strongest heterogeinity (SSH) at D20,
respectively.
The striking feature is that HDA, as an initial state,
and HDA', as an intermediate state, show a clear enhancement
at $Q\le 0.6$~\AA $^{-1}$.
This enhancement is absent in vHDA and LDA' and as
reported by us in references \cite{Schober-PB-1998,Schober-PRL-2000}
it is equally missing in LDA.
The low--$Q$ signal gives evidence for the heterogeneous
nature of the states that in terms of the wide-angle $S(Q)$
are intermediate to vHDA and LDA'.
This holds no matter how the intermediate modifications
were obtained in our experiments.

The enhanced signal in the low--$Q$ range coincides well
with the behaviour of the wide-angle $S(Q)$.
This is directly demonstrated by the broadening of the structure
factor maximum in the raw data (Fig.~\ref{fig_01}).
It can also be visualized by computing the Fourier transform 
$D(r)$ of $S(Q)$ shown in Fig.~\ref{fig_02}~c \cite{data_analysis}.
Obviously, in contrast to the properties of vHDA and LDA'
oscillations in $D(r)$ of SSH are strongly suppressed beyond
$r > 10$~\AA , indicating an apparent reduction of spatial
correlations in the intermediate states.
Please note, that variation of the signal at $Q < 0.65$~\AA $^{-1}$
have been suppressed for the computation of the data shown
in Fig.~\ref{fig_02}~c,
thus, leaving $D(r)$ unperturbed by the low--$Q$ regime.

\begin{figure}
\includegraphics[angle=0,width=150mm]{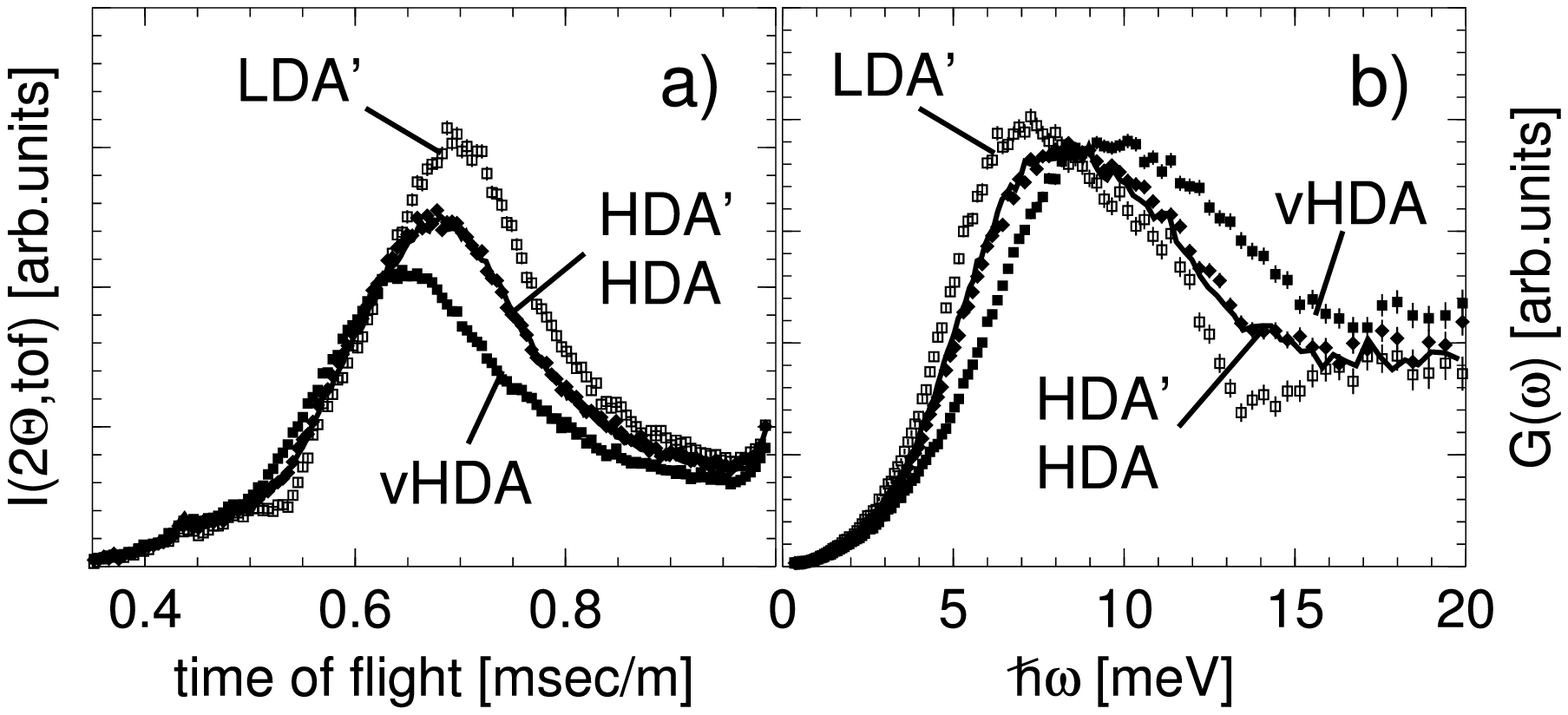}
\caption{(a) Inelastic intensity and (b) calculated
generalized density of states $G(\omega)$
measured at $T=75$~K at the spectrometer IN6.
}
\label{fig_03}
\end{figure}

The one--to--one correspondance of vHDA--to--LDA' and HDA--to--LDA
states does not only hold for the elastic signal but
is equally observed for molecular vibrations.
In Fig.~\ref{fig_03}~a we report the direct time--of--flight
signal $I(2\Theta,{\rm tof})$ measured at 75~K with sample \#\,1 at IN6
and averaged over the $2\Theta$ range sampled by the spectrometer.
Fig.~\ref{fig_03}~b shows the corresponding
generalized density of states $G(\omega)$.
It is obvious that no significant difference can be detected
in the spectra of HDA and HDA'.
The characteristic maximum in $G(\omega)$ which is due to
low-energy optic modes in the crystalline counterparts
shifts towards lower energies when vHDA transforms into
LDA' and the sample density decreases,
a feature as well observed for the diversity of crystalline
phases \cite{Li-JCP-1996,Koza-PRB-2004}.
Within the accuracy of the experiments the $G(\omega)$
of vHDA does not show any fingerprints of H--bond penetration,
whose effect on the inelastic response should result
in a pronounced splitting of the strong maximum.

\begin{figure}
\includegraphics[angle=0,width=150mm]{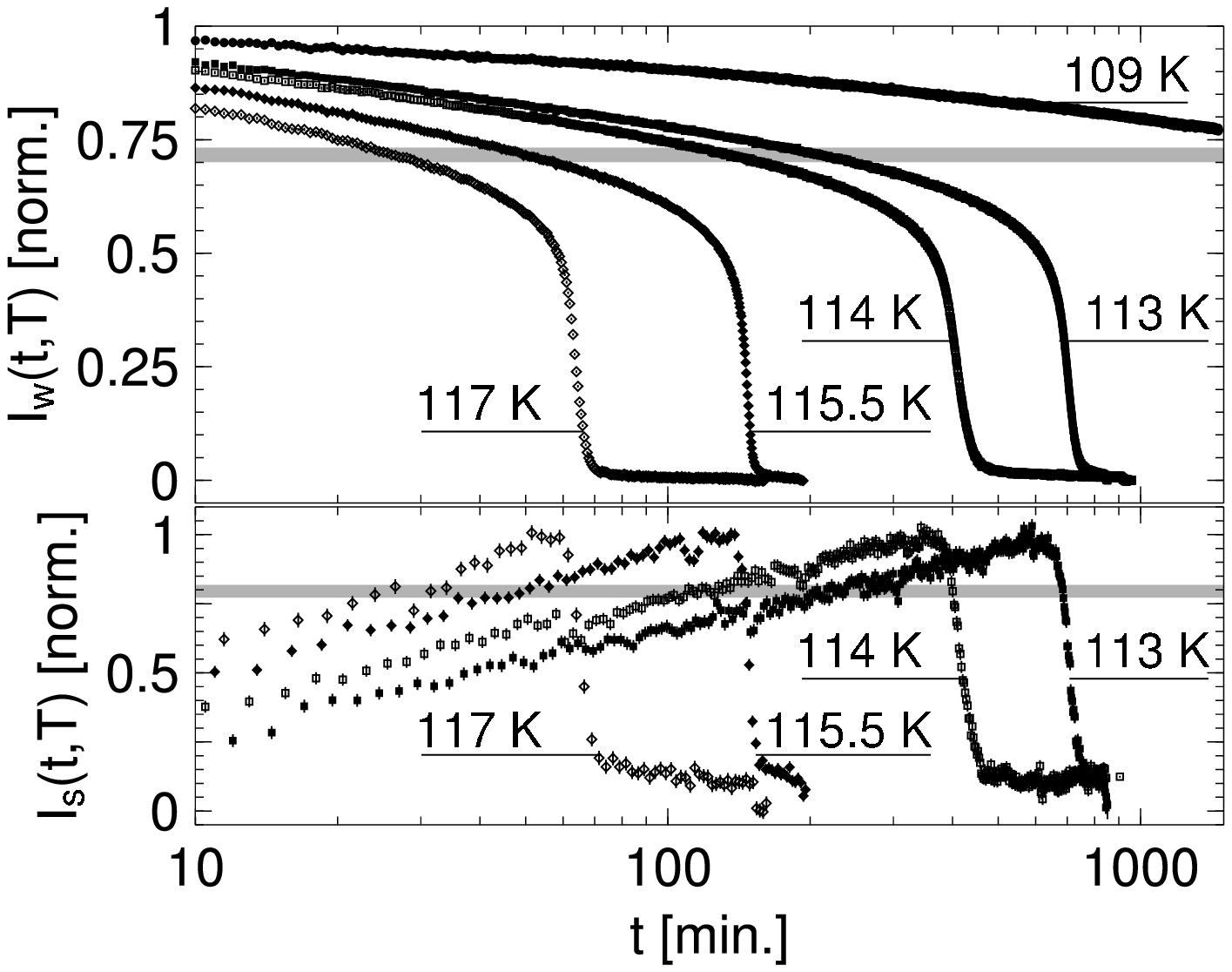}
\caption{On top, kinetics of the vHDA into LDA' transition
determined with sample \#\,2 shown in Fig.~\ref{fig_01}.
The nominal temperatures are given in the figure.
The grey shaded area indicates the position of
HDA' and respectively HDA in the plot.
At bottom, kinetics of the transient intensity
at low--$Q$ shown in Fig.~\ref{fig_02}~b.}
\label{fig_04}
\end{figure}

The fact that HDA and HDA' are identical in terms of $S(Q)$
and $G(\omega)$ poses the question of their alikeness in
the framework of thermodynamics.
To explore this point {\it in situ} experiments
of the transition vHDA to LDA' have been performed
on sample \#\,2 (5 portions of $\sim$0.5~ml each) at
temperatures 109~K, 113~K, 114~K, 115.5~K and 117~K.
Fig.~\ref{fig_04} shows the evolution of $S(Q)$
in the wide angle $I_{\rm w}(t,T)$ (top figure)
and small angle $I_{\rm s}(t,T)$ (bottom figure)
regime \cite{comment-vhda-2}.
By definition, $I_{\rm w}(t,T)$ equals 1 for vHDA and zero for LDA'.
Indicated as the grey shaded area is the stage of the transition
at which HDA' is identified.
It becomes immediately clear that at a given temperature 
vHDA transforms into LDA' on a longer time scale 
than HDA converts into LDA.
The characteristic time constants $\tau(T)$ of the transition
follow an Arrhenius line with an activation energy
$\Delta E \approx 65$~kJ/mol, i.e., a $\Delta E$ of
at least 20~kJ/mol higher than for the HDA to LDA transition
\cite{Koza-JPCM-2003}.
In other words, HDA' is structurally stable
in the temperature range in which HDA transforms
rapidly to LDA.

By combining a logarithmic term, which takes care of aging
\cite{Karpov-PRB-1993},
with an Avrami--Kolmogorov expression, which describes a first-order
transition, it is possible to reproduce $I(T,t)$ analytically
for the HDA to LDA transition \cite{Koza-JPCM-2003}.
This is not the case when starting the annealing process from vHDA.
As can be seen from fig.\ref{fig_04} the transition from HDA' to LDA'
becomes nearly uncontrollable at any temperature once it has set in.
Such a rapid transition kinetics is {\it a priori} incompatible with
the higher activation energy found for $\tau(T)$ when the system
is in equilibrium.
Therefore, an additional energy scale must be at work.
Given the speed of the transition the evacuation of latent heat
from the sample and thus the temperature control becomes a real
experimental challenge.

Indeed, following the vHDA to LDA' transition in high--vacuum
the vHDA samples recrystallize directly.
In contrast, LDA can be at any pressure obtained from HDA.
This behaviour can be easily understood by comparing the activation
energies of the vHDA to LDA' transition with the  activation
energy $\approx 66$~kJ/mol of recrystallisation
of hyper--quenched glassy water \cite{Hage-JCP-1995}.
This shows that the vHDA to LDA' transition energy is very
close to the recrystallisation limit of low--density
amorphous ice modifications.

In summary, we have shown that HDA, produced by compression
of hexagonal ice I$_{\rm h}$, and HDA', obtained as an intermediate
phase in the course of the transition vHDA to LDA', give indiscernible
responses in elastic and inelastic neutron scattering experiments.
A pronounced and transient low--$Q$ signal gives evidence for
heterogeneity in both HDA and HDA'.
They thus are similar beyond the local structural level.
Despite the fact that HDA and HDA' are indiscernible
in terms of their $S(Q)$ and $G(\omega)$ they are energetically
not identical as shown by the differences in transition kinetics.
It is not possible to tell from our experiments what properties
are responsible for this energetic difference.
However, it is reasonable to conjecture that the inequality holds 
as well for the final transition states LDA and LDA'
\cite{Giovambattista-PRL-2003,Guillot-JCP-2003}.

The resemblance of the HDA and HDA' response functions
shows unequivocally that static experiments probing local properties
in reciprocal space, here in particular wide angle diffraction, 
are insufficient to fully characterize the non--ergodic system.
The existence of a transient low--$Q$ diffraction signal demonstrates
the importance of collecting information in a region of reciprocal
space as large as possible to decide upon the spatial homogeneity
of the samples.
This does not hold only for experiments applying temperature
as a control parameter, but includes any study using
some external force, e.g. pressure \cite{Karpov-PRB-1993},
to which the non--ergodic sample is forced to respond.
Resemblance in terms of homogeneity still does not mean that
we deal physically with the same system.
Only time depending experiments give us information
on the energy states explored by the system 
\cite{Karpov-PRB-1993,Sciortino-JPCM-2001,Giovambattista-PRE-2004}.
In fact, HDA and HDA' are unequivocally different in this respect.

The present findings do in no way question the two--liquid scenario.
The seemingly homogeneous structures vHDA and LDA' are good candidates
for the glassy counterparts of the thermodynamic liquid states 
\cite{Poole-PRL-1994}.
Concerning local structure both HDA' and HDA belong to the vHDA
basin of states.
The transition from vHDA to LDA' involves heterogeneous intermediate
stages among which we find HDA'.
It bears all signs of a phase transition including the abrupt
nearly singular change of local structural order between HDA' and LDA'.
No other transitions or transition stages justifying the presence
of more than the two homogeneous structures could have been identified
\cite{Brovchenko-JCP-2003}.
As far as the properties of the wide--angle $S(Q)$ reported
here and in reference \cite{Koza-JPCM-2003} are concerned they
are in agreement with recently published results from molecular
dynamics simulations \cite{Guillot-JCP-2003,Martonak-PRL-2004}.
A computation of the properties observed in $S(Q)$ in the low--$Q$ range,
however require a larger simulation box size than the one used.

\bibliography{water_01}

\begin{thebibliography}{24}
\bibitem[{dat({\natexlab{a}})}]{data_analysis_01}
\bibinfo{note}{$S(Q)$ from IN6 data is obtained as:\\ $S(Q) =
  \int_{-\infty}^{+\infty} S(Q,\omega){\rm d}\omega$.}
\bibitem[{dat({\natexlab{b}})}]{data_analysis}
\bibinfo{note}{$D(r) \equiv \int_{0}^{\infty}(S(Q) - S(Q_{\rm
  max}))Q\sin{(Q\cdot r)}\rm{d}Q$, with $Q_{\rm max}$ the maximum accessed $Q$
  number. Spurious Bragg--peaks and arteficial detector spikes have been
  suppressed in $S(Q)$ data.}
\end{thebibliography}

\newpage


\end{document}